\documentclass[%
 reprint,
superscriptaddress,
 amsmath,amssymb,
 aps,
prb,
]{revtex4-2}

\newcommand{\Ea}{\ensuremath{{\cal E}_1}}
\newcommand{\Eb}{\ensuremath{{\cal E}_2}}
\newcommand{\Ec}{\ensuremath{{\cal E}_3}}

\newcommand{\Er}{\ensuremath{{\cal E}_{\rm R}}}
\newcommand{\Eo}{\ensuremath{{\cal E}_{1,2,3}}}

\usepackage{graphicx}
\usepackage{dcolumn}
\usepackage{bm}

\begin{document}

\preprint{APS/123-QED}

\title{Coherent dynamics of individual excitons in a quantum dot embedded in a nanopost}

\author{M.~Gaignard}%
\affiliation{Université Grenoble Alpes, CNRS, Grenoble INP, Institut Néel, 38000 Grenoble, France}

\author{K.~E.~Połczyńska}%
\affiliation{Institute of Experimental Physics, Faculty of Physics, University of Warsaw, ul. Pasteura 5, 02-093 Warsaw, Poland}

\author{N. Gregersen}%
\affiliation{DTU Electro, Department of Electrical and Photonics Engineering, Technical University of Denmark, DK-2800 Kongens Lyngby, Denmark}

\author{D. Wigger}%
\affiliation{Department of Physics, University of M\"{u}nster, 48149 M\"{u}nster, Germany}

\author{J.-P.~Poizat}
\affiliation{Université Grenoble Alpes, CNRS, Grenoble INP, Institut Néel, 38000 Grenoble, France}

\author{J.-M.~Gérard}
\affiliation{CEA, Univ. Grenoble Alpes, IRIG-PHELIQS, "Nanophysique et semiconducteurs" group, France}

\author{J.~Claudon}
\affiliation{CEA, Univ. Grenoble Alpes, IRIG-PHELIQS, "Nanophysique et semiconducteurs" group, France}

\author{J.~Kasprzak}
\email{jacek.kasprzak@cnrs.fr}
\affiliation{Université Grenoble Alpes, CNRS, Grenoble INP, Institut Néel, 38000 Grenoble, France}

\affiliation{Institute of Experimental Physics, Faculty of Physics, University of Warsaw, ul. Pasteura 5, 02-093 Warsaw, Poland}

\affiliation{Japanese-French lAboratory for Semiconductor physics and Technology (J-FAST), CNRS–Université Grenoble Alpes–Grenoble INP–University of Tsukuba, 1-1-1 Tennoudai, Tsukuba, 305-8573, Japan}

\date{\today}

\begin{abstract}
We measured coherent ultrafast dynamics of exciton complexes in a single strongly-confined InAs quantum dot embedded in a GaAs nanopost. Such a photonic structure combines a wave guiding with a cavity effect and assures an enhanced light-matter coupling over a broad spectral range of 30\,nm. Coherence properties of an exciton-biexciton system hosted by a quantum dot are assessed with four-wave mixing microscopy. Our results show that this broad-band photonic structure is an excellent asset to probe coherent couplings in a small set of solid state quantum systems and to investigate the coherence dynamics within the level structure of their excited states.     
\end{abstract}

\maketitle

\section{Introduction}

Photonic quantum technologies seek for bright single-photon sources, with a spectral purity and stability limited by their radiative broadening, providing a train of indistinguishable photons\,\cite{SantoriN02, Mosley2008, Alqedra2025}, required to generate entanglement\,\cite{Silberhorn2001, Versteegh2014, Schimpf2021}, to realize boson sampling\,\cite{Wang2017, Meer2020} and to achieve new frontiers in fundamental quantum optics and opto-mechanics\,\cite{Lagoudakis2016, Spinnler2024, Spinnler2024a, Gangopadhyay2025}. A reliable assessment of quantum light is provided by measuring its coherence, reflecting the capability of generating quantum superposition states.

\begin{figure*}[ht]
    \centering
\includegraphics[width=2.02\columnwidth]{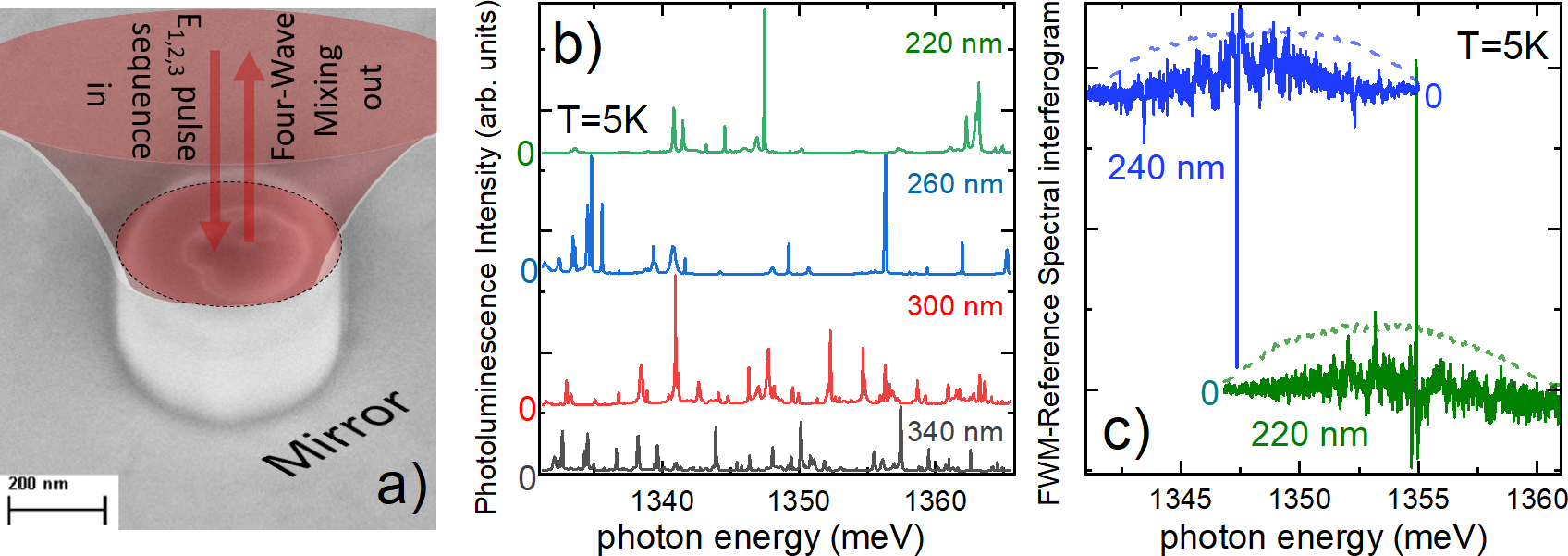}
    \caption{{\bf Photoluminescence and four-wave mixing response of individual InAs quantum dots embedded in nanoposts.} a)\,An exemplary scanning electron microscopy image of a nanopost, accompanied by a sketch of the resonant optical driving with a pulse sequence produced by a femtosecond laser. b)\,Exemplary photoluminescence spectra, measured on nanoposts with diameters as indicated, showing the broadband emission characteristics of our devices. c)\,Exemplary four-wave mixing spectral interferograms, measured on nanoposts with diameters of 220 and 240 nm, indicating the broadband operation of these devices in optical absorption, $(\tau_{12},\,\tau_{23})=(0.1,\,0)\,$ps. The spectra of the reference beam, \Er, are given in blue and green dashed lines, respectively.}
    \label{fig:1}
\end{figure*}

 The first-order coherence, commonly retrieved from the temporal decay of the interferometric visibility fringes, is often underestimated, due to inhomogeneous broadening induced by the residual spectral wandering of an optical transition. To disentangle homogeneous and inhomogeneous broadening within the spectral line shape, one has to apply nonlinear spectroscopy. In particular, heterodyne detected four-wave mixing (FWM)\,\cite{LangbeinOL06, Groll2025} turned out to be an excellent tool to measure coherence of solid state single-photon sources, i.e. excitons confined in epitaxial quantum dots (QDs). 
 
 The most mature is nowadays the InAs/GaAs platform, in which the dephasing indeed approaches the radiative lifetime limit, $T_2=2T_1$\,\cite{FrasNatPhot16, Spinnler2024, Spinnler2024a}. The coherence retrieval efficiency from single QDs has been greatly enhanced by embedding them in photonic structures, reducing the resonant laser background, while improving the collection efficiency. The strategies employ: i)\,cavities in the weak-coupling regime, generating field amplification around the emitter, accelerating the emission rate via the Purcell effect, although at the expense of a narrow band width\,\cite{FrasNatPhot16, Wigger2018, Kasprzak2022}, ii)\,optical antennas and wave guiding effects\,\cite{Bulgarini2014, MermillodPRL16, JakubczykACSPhot16, Wigger2023}, offering broadband operation, yet virtually without acceleration of the emission rate.

 \begin{figure*}
    \centering
\includegraphics[width=2.02\columnwidth]{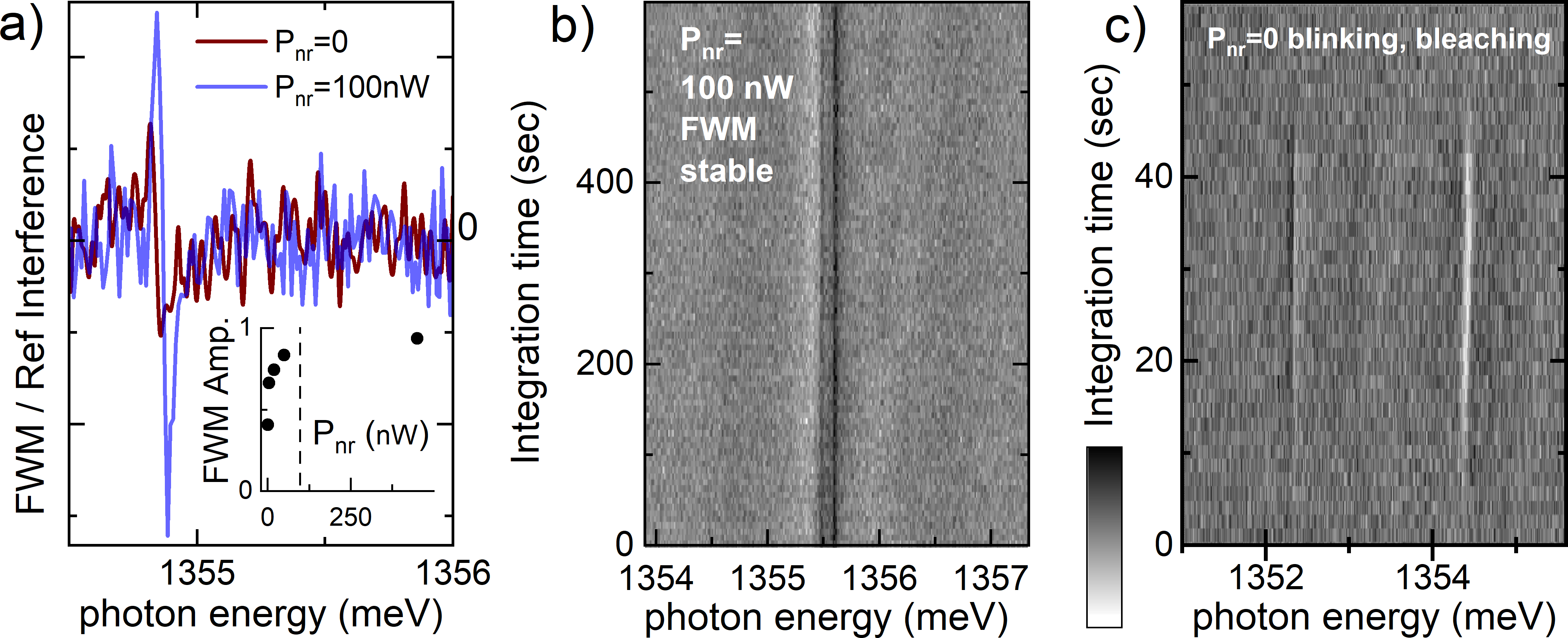}
    \caption{{\bf Influence of additional nonresonant illumination on coherent response of quantum dots in nanoposts.} a)\,FWM interferogram obtained within the same integration time and the same transition without  (brown) and with nonresonant excitation (blue). Inset: FWM amplitude versus intensity of the nonresonant excitation; P$_{nr}\simeq0.1\,\mu$W is sufficient to suppress the spectral wandering and  optimize the FWM signal. b)\,FWM interferogram versus integration time demonstrating a stabilized FWM interferogram. c)\,as b)\, without nonresonant illumination, exemplifying the QD blinking and bleaching in FWM, $(\tau_{12},\,\tau_{23})=(0.1,\,0)\,$ps in a),\,b)\,and c).}
    \label{fig:2}
\end{figure*}

\section{Sample}
It is desirable to combine the two, counteracting impacts, i.e., fully confine the light using a cavity to reach large Purcell factors, thus amplifying the source brightness, while maintaining broadband operation. A convenient solution has been recently provided by nanopost design and fabrication\,\cite{Kotal2021}. The structure is made of a trimmed GaAs nanowire of 500\,nm height, placed on top of a gold mirror that is coated with a 7\,nm-thick silica film, as shown by the scanning electron microscopy image in Fig.\,\ref{fig:1}\,a. This geometry accommodates three longitudinal anti-nodes of the electromagnetic field inside the body of the nanopost. Here, the quantum dot layer is located within the anti-node closest to the bottom. The diameter of the top-down fabricated nanoposts varies across the sample: we probe them from 150\,nm to 400\,nm in 10\,nm steps. However, the best results reported here were obtained using nanoposts with a diameter of 220\,nm, 230\,nm and 240\,nm. The details of the design, its optical performance and mode coupling versus geometrical parameters have been thoroughly discussed in Ref.\,\cite{Jacobsen2023}. In particular, the simulations therein indicate a Purcell bandwidth of 30\,nm full-width at half-maximum, as well as an ultra-broadband collection bandwidth exceeding 100\,nm, due to the inhibition of spontaneous emission into other modes, altogether showcasing the broadband operation of the nanopost.  

Fig.\,\ref{fig:1}\,b illustrates such broadband performance by showing typical photoluminescence spectra measured on several nanoposts with the indicated diameters. Indeed, an intense QD photoluminescence is measured over a broad spectral range of around 30\,meV. As the nanopost diameter decreases, the quantum dot photoluminescence intensity increases progressively, while the number of spectral lines reduces due to the reduced number of QDs inside the structure. The emission intensity is maximised for a nanopost diameter of around 220\,nm, with only a few transitions corresponding to the remaining emitters within the nanopost, presumably at its center. It is worth pointing out that, for this optimized diameter, the emission intensity of the dominant transitions is more than two orders of magnitude stronger than typical QD emission in planar structures. This dramatic increase in emission rate is evidence of the enhanced light-matter coupling engineered by these photonic structures. However, decreasing the nanopost diameter further results in line broadening and blinking due to increased spectral wandering, triggered by proximity to surface states. Ultimately, nanoposts with the smallest diameters exhibit virtually no emission due to the vanishingly small number of dots and the quenching of their optical activity. While this thorough photoluminescence characterisation indicates a broadband enhancement of the QD emission characteristics and increased out-coupling, it does not provide evidence of enhanced light in-coupling and the QD absorption. 

\section{Optimizing conditions for resonant four-wave mixing micro-spectroscopy}

Hence, in order to investigate (nonlinear) absorption of individual QDs in our nanoposts, heterodyne FWM microscopy is carried out.

In this experiment, we use three femtosecond laser pulses, \Eo, which are fully resonant with the targeted optical transition, as shown by the dashed lines in Fig.\,\ref{fig:1}\,c. The beams propagate in a collinear manner and are focused with a microscope objective to form a diffraction-limited spot, where $\Eo$ overlap spatially in full. Due to the intrinsic nonlinearity of the two-level system generated by the QD exciton\,\cite{Groll2025}, its coherent optical response can be induced, which, in the lowest order, is proportional to the third-order optical polarization $\mu^4\Ea^{\star}\Eb\Ec$, where $\mu$ denotes the transition dipole moment and $\star$ is phase conjugation. 

The challenge of this experiment lies in detecting such a defined FWM signal in the presence of a resonant background generated by \Eo, which is typically six orders of magnitude more dominant in the field. This is accomplished by: i)\,modulating the pulse trains with distinct radio frequencies in 100\,MHz range, $\Omega_{1,2,3}$, using acousto-optic deflection. Note that these RF frequency up-shifts correspond to just a few hundred nano-electronvolts, which introduces no detectable changes to the optical spectra. ii)\,demodulating (down-shifting) the reflectance containing a tiny amount of FWM at its corresponding heterodyne frequency, $\Omega_{3}+\Omega_{2}-\Omega_{1}$; and iii)\,interfering such a response with the reference laser, $\Er$, and subtracting the background in a balanced fashion. As such, by combining heterodyne detection with spectral interferometry, we achieve shot-noise-limited detection of the coherent response of individual QDs, which takes the form of spectrally resolved interferograms, as depicted in Fig.\,\ref{fig:1}\,c. Here, the FWM of individual QDs' zero-phonon lines takes form of sharp peaks building up at the corresponding transition energies. We present two examples measured on nanoposts with diameter of 220 and 240\,nm, which demonstrate that transitions separated by up to 12\,meV can be efficiently driven and detected. This proves that the nanopost geometry also assures broad-band performance in the optical absorption experiments. 

 When working with this first-generation of nanoposts containing InAs QDs, it has been established that the FWM spectroscopy can only be carried out in a presence of an additional, weak CW non-resonant optical pumping into a wetting layer. Such additional excitation does not play any role in the FWM detection scheme. Crucially though, it stabilizes charge fluctuations in the vicinity of the QD\,\cite{MajumdarPRB11, ArnoldPRX14} and at the surface\,\cite{HaPRB15, MannaASS20} of the nanopost, furthermore saturating the defects acting as traps for a non-radiative recombination. To this end, we employ a CW laser diode with a center energy of 1.49\,eV (830\,nm) with an average power of around 0.2\,$\mu$W. When attempting FWM spectroscopy with no such an extra illumination, we observe substantially larger spectral wandering of the excitonic transitions reaching several hundred $\mu$eV, as well as frequency drifts and absorption bleaching on a timescale from seconds to a couple of hours, as represented in Fig.\,\ref{fig:2}.    

Hence, with such optimized conditions (see inset in Fig.\,\ref{fig:2}\,a) for extra illumination, we achieve a FWM signal-to-noise ratio comparable with recently employed broad band photonic structures with individual InAs QDs: photonic waveguide antennas\,\cite{MermillodPRL16}, microlenses\,\cite{JakubczykACSPhot16}, and bullseye cavities\,\cite{Wigger2023}. As a FWM response from single, strongly-confined excitons has only been retrieved from a few types of photonic devices, it is beneficial to add the nanopost approach to this portfolio.

\begin{figure}[ht]
    \centering
\includegraphics[width=1.02\columnwidth]{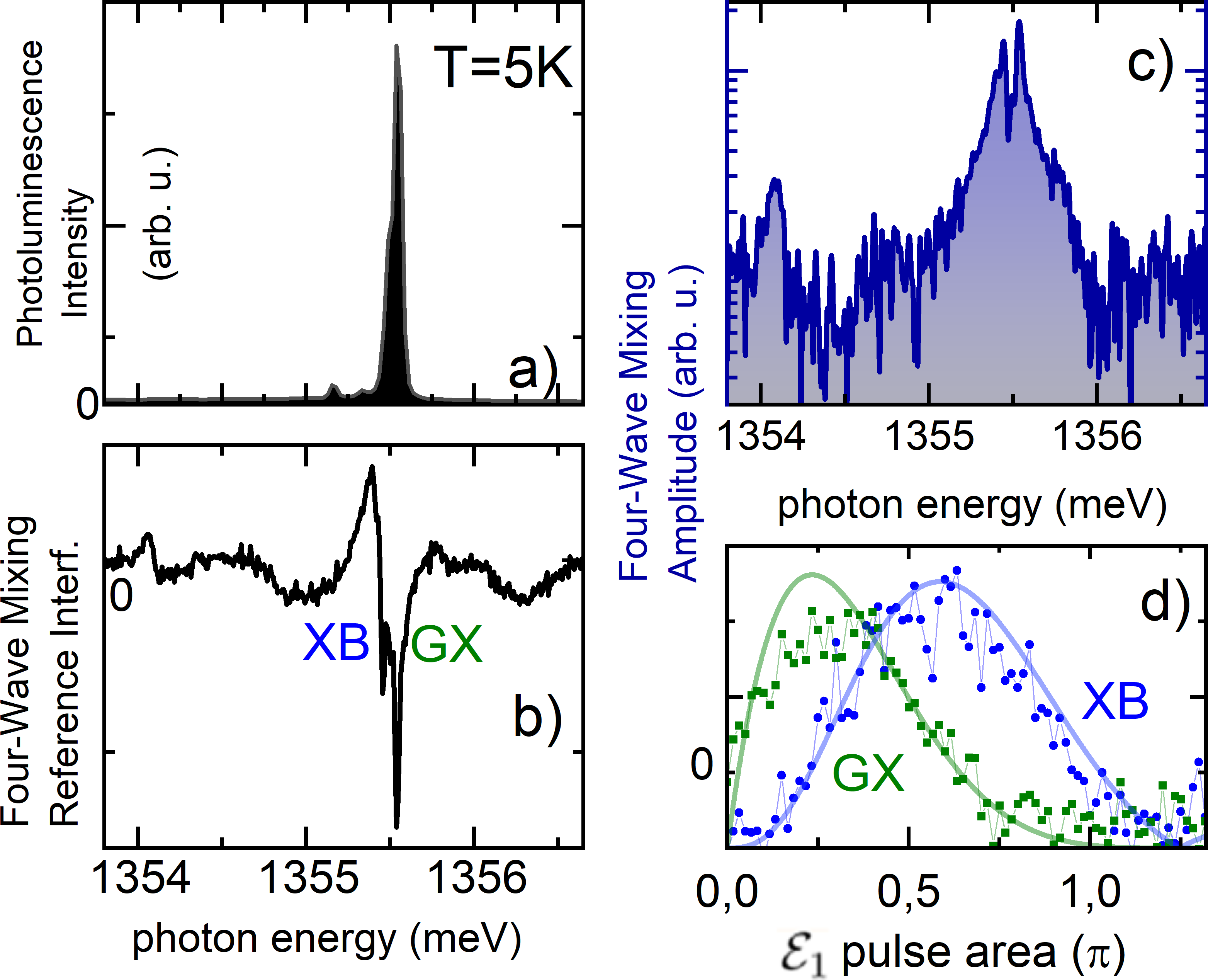}
    \caption{{\bf Photoluminescence and four-wave mixing response of a single InAs quantum dot embedded in a nanopost further investigated in this work.} a)\,The measured photoluminescence spectrum shows a record enhancement factor with respect to InAs quantum dots in planar structures. b)\,Spectral interference between the four-wave mixing field and the reference field $\Er$ measured under co-linear polarisation, showing the investigated exciton-biexciton (GX-XB) complex, $(\tau_{12},\,\tau_{23})=(0.1,\,0)\,$ps. An acoustic-phonon sideband around the zero-phonon line is also visible. c)\,FWM amplitude resulting from the interferogram shown in b), obtained by spectral interferometry. Note the logarithmic vertical scale. e)\,$\Ea$ pulse area dependence displays Rabi-like flopping of exciton and biexciton transitions. Simulations are given by the solid lines. The $\pi/2$ pulse area is attained for an $\Ea$ field of only $18.7\,\sqrt{\rm nW}$ indicating an excellent light-matter coupling.}
    \label{fig:3}
\end{figure}

\begin{figure}
    \centering
\includegraphics[width=1.01\columnwidth]{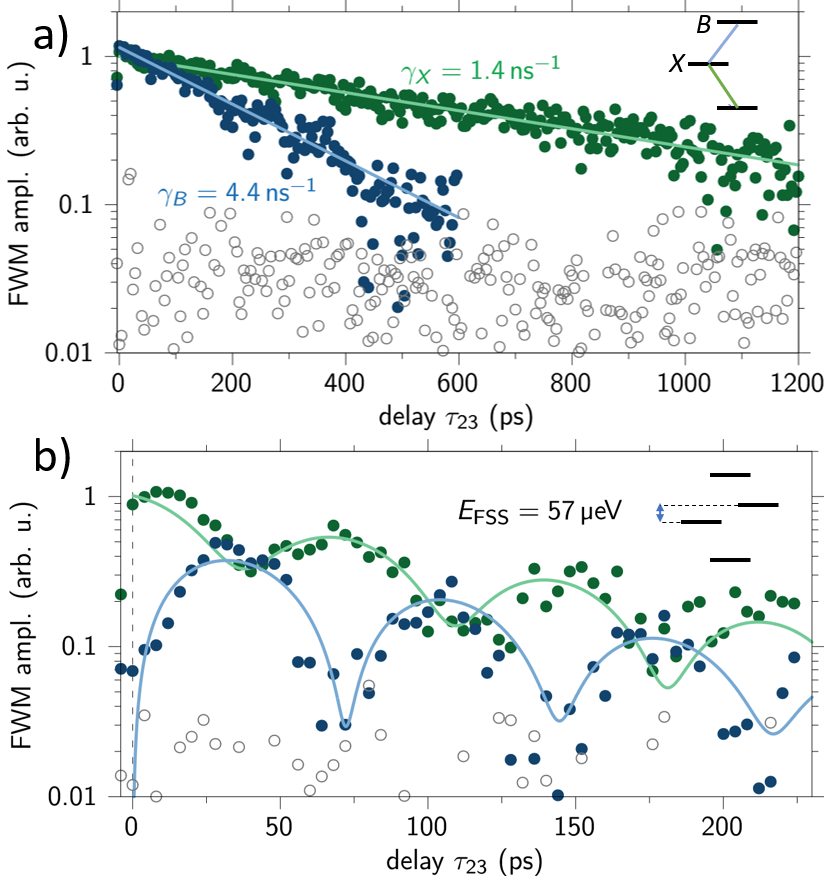}
    \caption{{\bf Density dynamics of the exciton-biexciton system of a single InAs quantum dot embedded in a nanopost measured via four-wave mixing.} a)\,Co-linear polarization excitation, showing an exponential decay, directly yielding exciton (X) and biexciton (B) population decay rates, $\tau_{12}=0\,$ps. b)\,Co-circular polarization excitation, showing a quantum beat resulting from the Raman coherence between fine-structure split states of the exciton, $\tau_{12}=0\,$ps. The modeling is given with solid green and blue lines, for GX and XB transitions, respectively. Modeling is given by green and blue traces, for GX and XB transitions respectively. Open circles indicate the noise level.}
    \label{fig:4}
\end{figure}

\section{Density and coherent dynamics of a single QD exciton in a nanopost measured with four-wave mixing}

In the following, we focus on the dominating excitonic complex, containing in fact four transitions, already visible in the spectral interferogram shown in Fig.\,\ref{fig:3}\,b. The oscillatory signal visible within a spectral window of several meV on both sides of the resonances is attributed to acoustic phonon sidebands\,\cite{Wigger2020} around the zero-phonon line. By performing FWM delay dynamics (see below) the complex is unequivocally attributed to a four-level exciton-biexciton system\,\cite{MermillodOptica16} with a biexciton binding energy of $\Delta$=+220\,$\mu$eV (biexciton transition on the lower energy side) and the excitons’ fine-structure splitting of $\delta$=57\,$\mu$eV. 

In Fig.\,\ref{fig:3}\,d, we show FWM amplitude as a function of the pulse area of the first driving beam (while maintaining the two other driving beams in a $\chi^{(3)}$ limit), for ground state-exciton (GX) and exciton–biexciton (XB) transitions. The data are fitted employing the Rabi rotation model of an exciton-biexciton system\,\cite{Wigger2017}. We point out that the $\pi$/2 pulse area is attained for a low external average intensity for the first driving beam $\Ea$ of barely $0.35\,\mu$W, with the pulse duration of $\simeq0.5\,$ps. This increased pulse duration is due to spectral shaping of $\Eo$ around the GX-XB complex for this Rabi-rotation experiment, increasing the effective pulse area. The small average power required to generate the $\pi$/2-pulse indicates an enhanced in-coupling of the laser into the nanopost and reaching the QD, altogether confirming  an excellent light-matter coupling realized by the nano-post structure.

\begin{figure}
    \centering
\includegraphics[width=1.02\columnwidth]{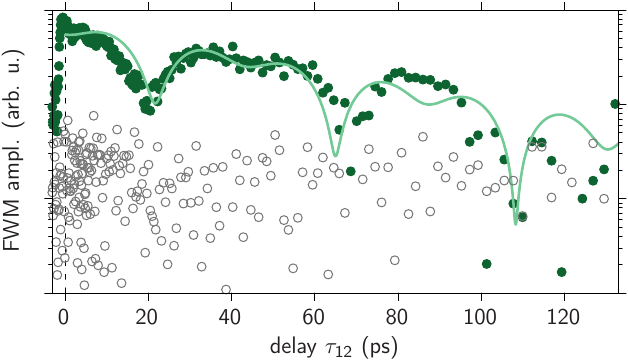}
    \caption{{\bf The coherence dynamics of the exciton confined in a single InAs quantum dot embedded in a nanopost.} An involved oscillatory structure is due to interplay of the fine-structure splitting and exciton-biexciton beating, strongly depending on the applied pulse areas, $\tau_{23}=0\,$ps. From the overall coherence decay we can estimate the dephasing time of T$_2 \simeq 240\,$ps, bearing into account the correction due to a finite spectral-temporal resolution of the spectrometer. The modeling is given by the solid green trace. Open circles indicate the noise level.}
    \label{fig:5}
\end{figure}

Until this point the driving beams $\Eo$ impinged the sample virtually at the same time. However, the full potential of the FWM spectroscopy is realized by scanning the respective time delays, $\tau_{12}$ and $\tau_{23}$, within the $\Eo$ pulse sequence:  

Namely, time-integrated FWM amplitude as a function of the delay between the last two arriving beams $\tau_{23}$ reflects the excitons’ population dynamics, and is presented in Fig.\,\ref{fig:4}. Under a co-linear polarization driving a), the GX and XB transition display exponential decays, as governed by the radiative rates of $\gamma_{{\rm X}}=1.4\,$ns$^{-1}$ and $\gamma_{{\rm B}}=4.4\,$ns$^{-1}$. The exciton radiative lifetime is shortened by a factor of two, with respect to QDs in a bulk environment\,\cite{BorriPRB02}. This points towards the Purcell factor of around 2, induced by the nanopost. Furthermore, note a shorter $T_1$ for the biexciton, as expected\,\cite{BacherPRL99, MermillodPRL16}, which in an ideal case should be a one-half of the GX transition's lifetime. 

Conversely, under co-circular driving, shown in b), the occupation dynamics displays a quantum beating. The beat originates from the Raman coherence between both fine-structure split excitons\,\cite{MermillodOptica16}, and is out of phase for GX and BX transitions. Due to this phase-shift, the excition and biexciton transitions can be readily recognized and separated in spectrally-resolved FWM, in spite of their spectral proximity.

Finally, the evolution of the exciton's coherence, as measured with the $\tau_{12}$-dependence, is presented in Fig.\,\ref{fig:5}. Access to the transition coherence dynamics via the $\tau_{12}$ sequence is unique to FWM spectroscopy and cannot be achieved using more common pump-probe experiments. Furthermore, due to the phase conjugation $(\star)$ between $\Ea$ and FWM, time-resolved FWM takes the form of a photon echo, enabling the homogeneous and inhomogeneous contributions to the spectral line shapes of the investigated transition to be accurately assessed\,\cite{Groll2025}. 

Coming back to the measured coherence dynamics in Fig.\,\ref{fig:5}, is actually quite involved, containing:
\begin{itemize}
    \item An overall (exponential) decay due to the dephasing of the zero phonon line, yielding the dephasing time $T_2\simeq240\,$ps, after correcting the measured dynamics with respect to the spectral resolution of the spectrometer\cite{JakubczykACSPhot16}.
    \item The FWM decay during the initial a few ps is attributed to the phonon-induced dephasing\,\cite{BorriPRB02, Wigger2017, Wigger2020} (note a vertical logarithmic scale).
    \item Finally, an involved oscillatory structure appears and is due to the fine-structure beating already observed in Fig.\,\ref{fig:3}\,b and a faster modulation stemming from the biexciton beating, with the contrast dependent on the applied pulse areas.
\end{itemize}

\section{Conclusions and outlook}
In conclusion, we exploited increased light-matter coupling in a nanopost photonic structure to efficiently measure the coherent nonlinear response of an exciton-biexciton system strongly confined in an InAs quantum dot. Using the FWM signal, we measured the coherence and population dynamics on a nanosecond timescale. With a Rabi rotation experiment we characterized the light-matter coupling by assigning pulse areas to the driving intensity. We emphasized the importance of additional weak non-resonant illumination in stabilizing the quantum dot's electronic environment, which substantially reduces spectral wandering and enhances the build-up of the heterodyne-detected FWM signal.

In the future, we plan to exploit the broadband operation of the nanopost to determine the level structure, dephasing and lifetime of QD-excited exciton states\,\cite{BennyPRB12}, as well as possible coherent couplings between these states, using two-dimensional spectroscopy. Another appealing prospect is to exploit the excellent optomechanical properties of these structures\,\cite{YeoNatNano14} to probe mutual interactions within hybrid quantum systems\,\cite{Kettler2020} using coherent nonlinear spectroscopy.

\section*{Acknowledgments}
We thank S. Kotal and A. Artioli for their contributions in sample fabrication. J.K. acknowledges the financial support from National Science Centre Poland, project no. 2023/51/B/ST3/01710.
 \bibliography{kasprzak39}

@ARTICLE{BacherPRL99,
  author = {G. Bacher and R. Weigand and J. Seufert and V. D. Kulakovskii and
	N. A. Gippius and A. Forchel and K. Leonardi and D. Hommel},
  title = {Biexciton versus Exciton Lifetime in a Single Semiconductor Quantum
	Dot},
  journal = {Phys. Rev. Lett.},
  year = {1999},
  volume = {83},
  pages = {4417}
}

@ARTICLE{BennyPRB12,
  author = {Benny, Y. and Kodriano, Y. and Poem, E. and Gershoni, D. and Truong,
	T. A. and Petroff, P. M.},
  title = {Excitation spectroscopy of single quantum dots at tunable positive,
	neutral, and negative charge states},
  journal = {Phys. Rev. B},
  year = {2012},
  volume = {86},
  pages = {085306},
  month = {Aug},
  issue = {8},
  numpages = {7},
  publisher = {American Physical Society}
}

@ARTICLE{BorriPRB02,
  author = {P. Borri and W. Langbein and S. Schneider and U. Woggon and R. L.
	Sellin and D. Ouyang and D. Bimberg},
  title = {Rabi Oscillations in the Excitonic Ground-State Transition of {InGaAs}
	Quantum Dots},
  journal = {Phys. Rev. B},
  year = {2002},
  volume = {66},
  pages = {081306(R)}
}

@Article{FrasNatPhot16,
  author    = {F. Fras and Q. Mermillod and G. Nogues and C. Hoarau and C. Schneider and M. Kamp and S. H\"{o}fling and W. Langbein and J. Kasprzak},
  journal   = {Nat. Phot.},
  title     = {Multi-Wave Coherent Control of a Solid State Single Emitter},
  year      = {2016},
  pages     = {155},
  volume    = {10},
}

@ARTICLE{JakubczykACSPhot16,
  author = {Tomasz Jakubczyk and Valentin Delmonte and Sarah Fischbach and Daniel
	Wigger and Doris E. Reiter and Quentin Mermillod and Peter Schnauber
	and Arsenty Kaganskiy and Jan-Hindrik Schulze and Andr\'{e} Strittmatter
	and Sven Rodt and Wolfgang Langbein and Tilmann Kuhn and Stephan
	Reitzenstein and Jacek Kasprzak},
  title = {Impact of Phonons on Dephasing of Individual Excitons in Deterministic
	Quantum Dot Microlenses},
  journal = {ACS Photonics},
  year = {2016},
  volume = {3},
  pages = {2461--2466}
}

@ARTICLE{LangbeinOL06,
  author = {W. Langbein and B. Patton},
  title = {Heterodyne Spectral Interferometry for Multidimensional Nonlinear
	Spectroscopy of Individual Quantum Systems},
  journal = {Opt. Lett.},
  year = {2006},
  volume = {31},
  pages = {1151},
  number = {8}
}

@ARTICLE{MermillodPRL16,
  author = {Mermillod, Q. and Jakubczyk, T. and Delmonte, V. and Delga, A. and
	Peinke, E. and G{\'e}rard, J.-M. and Claudon, J. and Kasprzak, J.},
  title = {Harvesting, Coupling, and Control of Single-Exciton Coherences in
	Photonic Waveguide Antennas},
  journal = {Phys. Rev. Lett.},
  year = {2016},
  volume = {116},
  pages = {163903}
}

@Article{MermillodOptica16,
  author    = {Mermillod, Q. and Wigger, D. and Delmonte, V. and Reiter, D. E. and Schneider, C. and Kamp, M. and H{\"o}fling, S. and Langbein, W. and Kuhn, T. and Nogues, G. and Kasprzak, J.},
  journal   = {Optica},
  title     = {Dynamics of excitons in individual {I}n{A}s quantum dots revealed in four-wave mixing spectroscopy},
  year      = {2016},
  pages     = {377},
  volume    = {3},
}

@Article{SantoriN02,
  author    = {Charles Santori and David Fattal and Jelena Vu\v{c}kovi\'{c} and Glen S.Solomon and Yoshihisa Yamamoto},
  journal   = {Nature},
  title     = {Indistinguishable Photons from a Single-Photon Device},
  year      = {2002},
  pages     = {594-597},
  volume    = {419},
}

@Article{YeoNatNano14,
  author    = {I. Yeo and P-L. de Assis and A. Gloppe and E. Dupont-Ferrier andP. Verlot and N. S. Malik and E. Dupuy and J. Claudon and J-M. G\'{e}rard and A. Auff\`{e}ves and G. Nogues and S. Seidelin and J-Ph. Poizat andO. Arcizet and M. Richard},
  journal   = {Nat. Nanotech.},
  title     = {Strain-mediated coupling in a quantum dot-mechanical oscillator hybrid system},
  year      = {2014},
  pages     = {106-110},
  volume    = {9},
}

@Article{Mosley2008,
  author    = {Mosley, Peter J. and Lundeen, Jeff S. and Smith, Brian J. and Wasylczyk, Piotr and U${'}$Ren, Alfred B. and Silberhorn, Christine and Walmsley, Ian A.},
  journal   = {Physical Review Letters},
  title     = {Heralded Generation of Ultrafast Single Photons in Pure Quantum States},
  year      = {2008},
  issn      = {1079-7114},
  month     = apr,
  number    = {13},
  pages     = {133601},
  volume    = {100},
  doi       = {10.1103/physrevlett.100.133601},
  publisher = {American Physical Society (APS)},
}

@Article{Silberhorn2001,
  author    = {Silberhorn, Ch. and Lam, P. K. and Wei{\ss}, O. and K\"{o}nig, F. and Korolkova, N. and Leuchs, G.},
  journal   = {Physical Review Letters},
  title     = {Generation of Continuous Variable {E}instein-{P}odolsky-{R}osen Entanglement via the {K}err Nonlinearity in an Optical Fiber},
  year      = {2001},
  issn      = {1079-7114},
  month     = may,
  number    = {19},
  pages     = {4267--4270},
  volume    = {86},
  doi       = {10.1103/physrevlett.86.4267},
  publisher = {American Physical Society (APS)},
}

@Article{Schimpf2021,
  author    = {Schimpf, Christian and Reindl, Marcus and Huber, Daniel and Lehner, Barbara and Covre Da Silva, Saimon F. and Manna, Santanu and Vyvlecka, Michal and Walther, Philip and Rastelli, Armando},
  journal   = {Science Advances},
  title     = {Quantum cryptography with highly entangled photons from semiconductor quantum dots},
  year      = {2021},
  issn      = {2375-2548},
  month     = apr,
  number    = {16},
  volume    = {7},
  doi       = {10.1126/sciadv.abe8905},
  publisher = {American Association for the Advancement of Science (AAAS)},
}

@Article{Wang2017,
  author    = {Wang, Hui and He, Yu and Li, Yu-Huai and Su, Zu-En and Li, Bo and Huang, He-Liang and Ding, Xing and Chen, Ming-Cheng and Liu, Chang and Qin, Jian and Li, Jin-Peng and He, Yu-Ming and Schneider, Christian and Kamp, Martin and Peng, Cheng-Zhi and H\"{o}fling, Sven and Lu, Chao-Yang and Pan, Jian-Wei},
  journal   = {Nature Photonics},
  title     = {High-efficiency multiphoton boson sampling},
  year      = {2017},
  issn      = {1749-4893},
  month     = may,
  number    = {6},
  pages     = {361--365},
  volume    = {11},
  doi       = {10.1038/nphoton.2017.63},
  publisher = {Springer Science and Business Media LLC},
}

@Article{Meer2020,
  author    = {van der Meer, R. and Renema, J. J. and Brecht, B. and Silberhorn, C. and Pinkse, P. W. H.},
  journal   = {Physical Review A},
  title     = {Optimizing spontaneous parametric down-conversion sources for boson sampling},
  year      = {2020},
  issn      = {2469-9934},
  month     = jun,
  number    = {6},
  pages     = {063821},
  volume    = {101},
  doi       = {10.1103/physreva.101.063821},
  publisher = {American Physical Society (APS)},
}

@Article{Kasprzak2022,
  author    = {Kasprzak, Jacek and Wigger, Daniel and Hahn, Thilo and Jakubczyk, Tomasz and Zinkiewicz, {\L}ukasz and Machnikowski, Pawe{\l} and Kuhn, Tilmann and Motte, Jean-FranÃ§ois and Pacuski, Wojciech},
  journal   = {ACS Photonics},
  title     = {Coherent Dynamics of a Single {M}n-Doped Quantum Dot Revealed by Four-Wave Mixing Spectroscopy},
  year      = {2022},
  issn      = {2330-4022},
  month     = feb,
  number    = {3},
  pages     = {1033--1041},
  volume    = {9},
  doi       = {10.1021/acsphotonics.1c01981},
  publisher = {American Chemical Society (ACS)},
}

@Article{Wigger2018,
  author    = {Wigger, Daniel and Schneider, Christian and Gerhardt, Stefan and Kamp, Martin and H\"{o}fling, Sven and Kuhn, Tilmann and Kasprzak, Jacek},
  journal   = {Optica},
  title     = {Rabi oscillations of a quantum dot exciton coupled to acoustic phonons: coherence and population readout},
  year      = {2018},
  issn      = {2334-2536},
  month     = nov,
  number    = {11},
  pages     = {1442},
  volume    = {5},
  doi       = {10.1364/optica.5.001442},
  publisher = {The Optical Society},
}

@Article{Wigger2023,
  author    = {Wigger, Daniel and Schall, Johannes and Deconinck, Marielle and Bart, Nikolai and Mrowi\'{n}ski, Pawe{\l} and Krzykowski, Mateusz and Gawarecki, Krzysztof and von Helversen, Martin and Schmidt, Ronny and Bremer, Lucas and Bopp, Frederik and Reuter, Dirk and Wieck, Andreas D. and Rodt, Sven and Renard, Julien and Nogues, Gilles and Ludwig, Arne and Machnikowski, PaweÂ³ and Finley, Jonathan J. and Reitzenstein, Stephan and Kasprzak, Jacek},
  journal   = {ACS Photonics},
  title     = {Controlled Coherent Coupling in a Quantum Dot Molecule Revealed by Ultrafast Four-Wave Mixing Spectroscopy},
  year      = {2023},
  issn      = {2330-4022},
  month     = may,
  number    = {5},
  pages     = {1504--1511},
  volume    = {10},
  doi       = {10.1021/acsphotonics.3c00108},
  publisher = {American Chemical Society (ACS)},
}

@Article{Kotal2021,
  author    = {Kotal, Saptarshi and Artioli, Alberto and Wang, Yujing and Osterkryger, Andreas Dyhl and Finazzer, Matteo and Fons, Romain and Genuist, Yann and Bleuse, Joel and G\'{e}rard, Jean-Michel and Gregersen, Niels and Claudon, Julien},
  journal   = {Applied Physics Letters},
  title     = {A nanowire optical nanocavity for broadband enhancement of spontaneous emission},
  year      = {2021},
  issn      = {1077-3118},
  month     = may,
  number    = {19},
  volume    = {118},
  doi       = {10.1063/5.0045834},
  publisher = {AIP Publishing},
}

@Article{Jacobsen2023,
  author    = {Jacobsen, Martin Arentoft and Wang, Yujing and Vannucci, Luca and Claudon, Julien and G\'{e}rard, Jean-Michel and Gregersen, Niels},
  journal   = {Nanoscale},
  title     = {Performance of the nanopost single-photon source: beyond the single-mode model},
  year      = {2023},
  issn      = {2040-3372},
  number    = {13},
  pages     = {6156--6169},
  volume    = {15},
  doi       = {10.1039/d2nr07132k},
  publisher = {Royal Society of Chemistry (RSC)},
}

@Article{Wigger2020,
  author    = {Wigger, Daniel and Karakhanyan, Vage and Schneider, Christian and Kamp, Martin and H\"{o}fling, Sven and Machnikowski, PaweÂ³ and Kuhn, Tilmann and Kasprzak, Jacek},
  journal   = {Optics Letters},
  title     = {Acoustic phonon sideband dynamics during polaron formation in a single quantum dot},
  year      = {2020},
  issn      = {1539-4794},
  month     = feb,
  number    = {4},
  pages     = {919},
  volume    = {45},
  doi       = {10.1364/ol.385602},
  publisher = {Optica Publishing Group},
}

@Article{Wigger2017,
  author    = {Wigger, D. and Mermillod, Q. and Jakubczyk, T. and Fras, F. and Le-Denmat, S. and Reiter, D. E. and H\"{o}fling, S. and Kamp, M. and Nogues, G. and Schneider, C. and Kuhn, T. and Kasprzak, J.},
  journal   = {Physical Review B},
  title     = {Exploring coherence of individual excitons in {I}n{A}s quantum dots embedded in natural photonic defects: Influence of the excitation intensity},
  year      = {2017},
  issn      = {2469-9969},
  month     = oct,
  number    = {16},
  pages     = {165311},
  volume    = {96},
  doi       = {10.1103/physrevb.96.165311},
  publisher = {American Physical Society (APS)},
}

@Article{MannaASS20,
  author    = {Manna, Santanu and Huang, Huiying and da Silva, Saimon Filipe Covre and Schimpf, Christian and Rota, Michele B. and Lehner, Barbara and Reindl, Marcus and Trotta, Rinaldo and Rastelli, Armando},
  journal   = {Applied Surface Science},
  title     = {Surface passivation and oxide encapsulation to improve optical properties of a single {G}a{A}s quantum dot close to the surface},
  year      = {2020},
  issn      = {0169-4332},
  month     = dec,
  pages     = {147360},
  volume    = {532},
  doi       = {10.1016/j.apsusc.2020.147360},
  publisher = {Elsevier BV},
}

@Article{MajumdarPRB11,
  author    = {Majumdar, Arka and Kim, Erik D. and Vu\v{c}kovi\'{c}, Jelena},
  journal   = {Physical Review B},
  title     = {Effect of photogenerated carriers on the spectral diffusion of a quantum dot coupled to a photonic crystal cavity},
  year      = {2011},
  issn      = {1550-235X},
  month     = nov,
  number    = {19},
  pages     = {195304},
  volume    = {84},
  doi       = {10.1103/physrevb.84.195304},
  publisher = {American Physical Society (APS)},
}

@Article{HaPRB15,
  author    = {Ha, Neul and Mano, Takaaki and Chou, Ying-Lin and Wu, Yu-Nien and Cheng, Shun-Jen and Bocquel, Juanita and Koenraad, Paul M. and Ohtake, Akihiro and Sakuma, Yoshiki and Sakoda, Kazuaki and Kuroda, Takashi},
  journal   = {Physical Review B},
  title     = {Size-dependent line broadening in the emission spectra of single {G}a{A}s quantum dots: Impact of surface charge on spectral diffusion},
  year      = {2015},
  issn      = {1550-235X},
  month     = aug,
  number    = {7},
  pages     = {075306},
  volume    = {92},
  doi       = {10.1103/physrevb.92.075306},
  publisher = {American Physical Society (APS)},
}

@Article{ArnoldPRX14,
  author    = {Arnold, C. and Loo, V. and Lemaitre, A. and Sagnes, I. and Krebs, O. and Voisin, P. and Senellart, P. and Lanco, L.},
  journal   = {Physical Review X},
  title     = {Cavity-Enhanced Real-Time Monitoring of Single-Charge Jumps at the Microsecond Time Scale},
  year      = {2014},
  issn      = {2160-3308},
  month     = apr,
  number    = {2},
  pages     = {021004},
  volume    = {4},
  doi       = {10.1103/physrevx.4.021004},
  publisher = {American Physical Society (APS)},
}

@Article{Groll2025,
  author    = {Groll, Daniel and Hahn, Thilo and Machnikowski, PaweÂ³ and Kuhn, Tilmann and Kasprzak, Jacek and Wigger, Daniel},
  journal   = {Nano Futures},
  title     = {Fundamentals of heterodyne wave mixing spectroscopy: a tutorial},
  year      = {2025},
  issn      = {2399-1984},
  month     = Nov,
  number    = {4},
  pages     = {042601},
  volume    = {9},
  doi       = {10.1088/2399-1984/ae108f},
  publisher = {IOP Publishing},
}

@Article{Spinnler2024,
  author    = {Spinnler, Clemens and Nguyen, Giang N. and Wang, Ying and Erbe, Marcel and Javadi, Alisa and Zhai, Liang and Scholz, Sven and Wieck, Andreas D. and Ludwig, Arne and Lodahl, Peter and Midolo, Leonardo and Warburton, Richard J.},
  journal   = {Physical Review Applied},
  title     = {Quantum dot coupled to a suspended-beam mechanical resonator: From the unresolved- to the resolved-sideband regime},
  year      = {2024},
  issn      = {2331-7019},
  month     = Mar,
  number    = {3},
  pages     = {034046},
  volume    = {21},
  doi       = {10.1103/physrevapplied.21.034046},
  publisher = {American Physical Society (APS)},
}

@Article{Spinnler2024a,
  author    = {Spinnler, Clemens and Nguyen, Giang N. and Wang, Ying and Zhai, Liang and Javadi, Alisa and Erbe, Marcel and Scholz, Sven and Wieck, Andreas D. and Ludwig, Arne and Lodahl, Peter and Midolo, Leonardo and Warburton, Richard J.},
  journal   = {Nature Communications},
  title     = {A single-photon emitter coupled to a phononic-crystal resonator in the resolved-sideband regime},
  year      = {2024},
  issn      = {2041-1723},
  month     = Nov,
  number    = {1},
  volume    = {15},
  doi       = {10.1038/s41467-024-53882-2},
  publisher = {Springer Science and Business Media LLC},
}

@Article{Kettler2020,
  author    = {Kettler, Jan and Vaish, Nitika and de Lépinay, Laure Mercier and Besga, Benjamin and de Assis, Pierre-Louis and Bourgeois, Olivier and Auff\`ees, Alexia and Richard, Maxime and Claudon, Julien and Gérard, Jean-Michel and Pigeau, Benjamin and Arcizet, Olivier and Verlot, Pierre and Poizat, Jean-Philippe},
  journal   = {Nature Nanotechnology},
  title     = {Inducing micromechanical motion by optical excitation of a single quantum dot},
  year      = {2020},
  issn      = {1748-3395},
  month     = Dec,
  number    = {3},
  pages     = {283--287},
  volume    = {16},
  doi       = {10.1038/s41565-020-00814-y},
  publisher = {Springer Science and Business Media LLC},
}

@Article{Versteegh2014,
  author    = {Versteegh, Marijn A. M. and Reimer, Michael E. and Jöns, Klaus D. and Dalacu, Dan and Poole, Philip J. and Gulinatti, Angelo and Giudice, Andrea and Zwiller, Val},
  journal   = {Nature Communications},
  title     = {Observation of strongly entangled photon pairs from a nanowire quantum dot},
  year      = {2014},
  issn      = {2041-1723},
  month     = Oct,
  number    = {1},
  volume    = {5},
  doi       = {10.1038/ncomms6298},
  publisher = {Springer Science and Business Media LLC},
}

@Article{Gangopadhyay2025,
  author    = {Gangopadhyay, Sayan and Yu, Lingxi and Patel, Tarun and Pennacchietti, Matteo and Northeast, David B. and Williams, Robin L. and Poole, Philip J. and Reimer, Michael E. and Dalacu, Dan},
  journal   = {Nano Letters},
  title     = {Mode-Matched Resonant Excitation of a Nanowire Quantum Dot in a Nanophotonic Waveguide},
  year      = {2025},
  issn      = {1530-6992},
  month     = Nov,
  number    = {49},
  pages     = {17096--17103},
  volume    = {25},
  doi       = {10.1021/acs.nanolett.5c04530},
  publisher = {American Chemical Society (ACS)},
}

@Article{Lagoudakis2016,
  author    = {Lagoudakis, K. G. and McMahon, P. L. and Dory, C. and Fischer, K. A. and Müller, K. and Borish, V. and Dalacu, D. and Poole, P. J. and Reimer, M. E. and Zwiller, V. and Yamamoto, Y. and Vu\v{c}kovi\'{c}, J.},
  journal   = {Optica},
  title     = {Ultrafast coherent manipulation of trions in site-controlled nanowire quantum dots},
  year      = {2016},
  issn      = {2334-2536},
  month     = Nov,
  number    = {12},
  pages     = {1430},
  volume    = {3},
  doi       = {10.1364/optica.3.001430},
  publisher = {Optica Publishing Group},
}

@Article{Alqedra2025,
  author    = {Alqedra, Mohammed K. and Huang, Chiao-Tzu and Chang, Wen-Hao and Haffouz, Sofiane and Poole, Philip J. and Dalacu, Dan and Elshaari, Ali W. and Zwiller, Val},
  journal   = {Applied Physics Letters},
  title     = {Indistinguishable single photons from nanowire quantum dots in the telecom O-band},
  year      = {2025},
  issn      = {1077-3118},
  month     = Oct,
  number    = {16},
  volume    = {127},
  doi       = {10.1063/5.0271146},
  publisher = {AIP Publishing},
}

@Article{Bulgarini2014,
  author    = {Bulgarini, Gabriele and Reimer, Michael E. and Bouwes Bavinck, Maaike and Jöns, Klaus D. and Dalacu, Dan and Poole, Philip J. and Bakkers, Erik P. A. M. and Zwiller, Val},
  journal   = {Nano Letters},
  title     = {Nanowire Waveguides Launching Single Photons in a Gaussian Mode for Ideal Fiber Coupling},
  year      = {2014},
  issn      = {1530-6992},
  month     = June,
  number    = {7},
  pages     = {4102--4106},
  volume    = {14},
  doi       = {10.1021/nl501648f},
  publisher = {American Chemical Society (ACS)},
}
\end{document}